\documentclass[aps,prd,nofootinbib,twocolumn,superscriptaddress,preprintnumbers,balancelastpage,longbibliography]{revtex4-1}
\usepackage{mwe} 
\usepackage{bm}
\usepackage{booktabs}
\usepackage{siunitx}
\usepackage{placeins}
\usepackage{amsmath,amssymb,mathtools,bm}
\usepackage{graphicx, color, hepunits}
\usepackage[dvipsnames]{xcolor}
\usepackage{cancel}
\usepackage[normalem]{ulem}
\usepackage{float}
\usepackage{multirow}
\usepackage{comment}
\usepackage{soul}
\usepackage{hyperref} 
\hypersetup{
    colorlinks=true,       
    linkcolor=blue,        
    citecolor=blue,        
    filecolor=magenta,     
    urlcolor=blue          
}
\usepackage[utf8]{inputenc}
\usepackage[english]{babel}

\newcommand{\tuple}{{\it{\Upsilon}}}

\makeatletter
\newcounter{savesection}
\newcounter{apdxsection}
\renewcommand\appendix{\par
  \setcounter{savesection}{\value{section}}%
  \setcounter{section}{\value{apdxsection}}%
  \setcounter{subsection}{0}%
  \gdef\thesection{\@Alph\c@section}}
\newcommand\unappendix{\par
  \setcounter{apdxsection}{\value{section}}%
  \setcounter{section}{\value{savesection}}%
  \setcounter{subsection}{0}%
  \gdef\thesection{\@arabic\c@section}}
\makeatother

\begin{document}

\title{Discovering $\mu$Hz gravitational waves and ultra-light dark matter\\
with binary resonances}
\author{Joshua~W.~Foster}
\affiliation{Astrophysics Theory Department, Theory Division, Fermilab, Batavia, IL 60510, USA}
\affiliation{Kavli Institute for Cosmological Physics, University of Chicago, Chicago, IL 60637, USA}
\email{jwfoster@wisc.edu}
\author{Diego~Blas}
\affiliation{Institut de F\'{i}sica d’Altes Energies (IFAE), The Barcelona Institute of Science and Technology,
Campus UAB, 08193 Bellaterra (Barcelona), Spain}
\affiliation{Instituci\'{o} Catalana de Recerca i Estudis Avan\c{c}ats (ICREA), Passeig Llu\'{i}s Companys 23, 08010 Barcelona, Spain}
\author{Adrien~Bourgoin}
\affiliation{LTE, Observatoire de Paris, Universit\'e PSL, Sorbonne Universit\'e, Universit\'e de Lille, LNE, CNRS 61 Avenue de l'Observatoire, 75014 Paris, France}
\author{Aurelien~Hees}
\affiliation{LTE, Observatoire de Paris, Universit\'e PSL, Sorbonne Universit\'e, Universit\'e de Lille, LNE, CNRS 61 Avenue de l'Observatoire, 75014 Paris, France}
\author{M\'iriam~Herrero-Valea}
\affiliation{Institut de F\'{i}sica d’Altes Energies (IFAE), The Barcelona Institute of Science and Technology,
Campus UAB, 08193 Bellaterra (Barcelona), Spain}
\affiliation{Barcelona Supercomputing Center (BSC), Plaça d'Eusebi Güell 1-3, 08034 Barcelona, Spain}
\author{Alexander~C.~Jenkins}
\affiliation{Kavli Institute for Cosmology, University of Cambridge, Madingley Road, Cambridge CB3 0HA, UK}
\affiliation{DAMTP, University of Cambridge, Wilberforce Road, Cambridge CB3 0WA, UK}
\author{Xiao~Xue}
\affiliation{Institut de F\'{i}sica d’Altes Energies (IFAE), The Barcelona Institute of Science and Technology,
Campus UAB, 08193 Bellaterra (Barcelona), Spain}

\preprint{FERMILAB-PUB-25-0091-T}

\date{\today}
\begin{abstract}
In the presence of a weak gravitational wave (GW) background, astrophysical binary systems act as high-quality resonators, with efficient transfer of energy and momentum between the orbit and a harmonic GW leading to potentially detectable orbital perturbations. In this work, we develop and apply a novel modeling and analysis framework that describes the imprints of GWs on binary systems in a fully time-resolved manner to study the sensitivity of lunar laser ranging, satellite laser ranging, and pulsar timing to both resonant and nonresonant GW backgrounds. We demonstrate that optimal data collection, modeling, and analysis lead to projected sensitivities which are orders of magnitude stronger than previously appreciated, opening up a new possibility for probing the physics-rich but notoriously challenging-to-access $\mu\mathrm{Hz}$ frequency band. This improvement arises from retaining accumulated orbital dephasing in the measured residual and scales quadratically with the ratio of the observing time to the unperturbed binary period, up to eccentricity-dependent factors. We also discuss improved prospects for the detection of the stochastic fluctuations of ultra-light dark matter, which may analogously perturb the binary orbits. 
\end{abstract}
\maketitle


The first detection of gravitational waves (GWs) by LIGO/Virgo/KAGRA~\cite{LIGOScientific:2016aoc} kicked off an exciting new era of GW astronomy that has the potential to provide transformational insights into aspects of fundamental particle physics, inflation, early Universe cosmology, black holes, and stellar astrophysics~\cite{Sathyaprakash:2009xs}. 
Reaching the full promise of GW science requires a comprehensive detection program that spans the many orders of magnitude of possible GW frequencies. 
Notably, GWs at frequencies between $10^{-7}$~Hz and $10^{-4}$~Hz --- a range above those accessible to pulsar timing arrays but below those accessible to LISA --- are particularly phenomenologically compelling, but observationally challenging.
This ``$\mu$Hz gap" in GW sensitivities has motivated several proposals, including $\mu\mathrm{ARES}$~\cite{Sesana:2019vho}, FRB timing~\cite{Lu:2024yuo}, astrometry with Gaia~\cite{Caliskan:2023cqm} or the Roman Space Telescope~\cite{Wang:2022sxn}, Doppler tracking of a Uranus orbiter~\cite{Zwick:2024hag}, and asteroid laser ranging~\cite{Fedderke:2021kuy}, typically requiring advances well beyond state-of-the-art to achieve desired sensitivity levels.

Here, we consider a different possibility for $\mu\mathrm{Hz}$ GW detection first studied in~\cite{Hui:2012yp, Blas:2021mpc, Blas:2021mqw} and achievable with current or near-future technology and datasets: that the resonant perturbations of binary systems by GWs could lead to deviations of celestial bodies from their unperturbed orbits that may be detected with precise tracking. 

A weak GW of large wavelength with respect to the size of the system acts on a binary system as a perturbing acceleration~\cite{Misner:1974qy}
\begin{equation}
    \bm{a}^{i} = \frac{1}{2}\ddot h_{ij}(t)\bm{r}^j,
    \label{eq:GWPerturbation}
\end{equation}
where $\bm{r}$ is the displacement vector between the orbiting bodies, and $h_{ij}(t)$ is the transverse-traceless part of the metric perturbation at the binary's center of mass. 
Resonance is achieved when the binary system is perturbed by a force that is harmonic with the unperturbed orbital period, leading to an efficient transfer of energy between the GW field and the binary system. As a result, the orbital motion is modified. 
These orbital perturbations accumulate in time, manifesting as, \textit{e.g.}, a time-evolving orbit period as studied\footnote{A related phenomenon of resonant locking with chirping GW signals that may occur for certain binary systems was presented in \cite{Sharpe:2025voe}, but this effect requires considerably stronger GW backgrounds than those to which we find we have sensitivity in this work.} in~\cite{Blas:2021mpc,Blas:2021mqw}. 
These works suggested that leading $\mu$Hz GW sensitivity through binary resonances may be within reach of existing lunar/satellite laser ranging and pulsar-timing data.

In a companion paper~\cite{Foster:2025csl}, we develop a new formalism for fully resolved predictions for the real-time dynamics of binary systems perturbed by GWs, finding the resulting observable perturbations to be \textit{orders of magnitude larger than previously estimated}. Based on these findings, in this \textit{Letter}, we develop new sensitivity projections for stochastic GW backgrounds (SGWBs) that make use of a broad array of astrophysical binary systems that can be measured and modeled with exquisite precision. As we demonstrate here, presently available data are likely sufficient to achieve leading sensitivity to $\mu$Hz GWs, and may even reach a strain sensitivity that tests well-motivated astrophysical or `new physics' mechanisms for generating GWs, with particular relevance for the recently claimed detection of a nHz SGWB by pulsar timing arrays~\cite{NANOGrav:2023hde, NANOGrav:2023gor}. We have also identified high-eccentricity orbits as those most promising for GW detection, suggesting future prospects for even deeper reach in GW strain through optimal designs for future laser ranging missions that we consider here.

Moreover, a resonance can be realized with any external acceleration that oscillates at frequencies harmonic with the orbital period.
One possible source of these external perturbations, aside from GWs, is dark matter of ultra-light mass $m_{\rm DM}$ (ultra-light dark matter, ULDM), which is one of the most actively explored models of current dark matter research~\cite{Ferreira:2020fam,Hui:2016ltb}. The nearly monochromatic fluctuations of cold ULDM result in fluctuations in the gravitational potential $\psi$, at frequencies $\omega\approx 2 m_{\rm DM}$ and $\omega\lesssim  10^{-6} m_{\rm DM}$~\cite{Bar-Or:2018pxz,Church:2018sro,Ferreira:2020fam,Kim:2023pkx}, 
causing a perturbing oscillating acceleration analogous to Eq.~\eqref{eq:GWPerturbation}~\cite{Hui:2012yp,Blas:2016ddr, Blas:2019hxz, Rozner:2019gba,Kus:2024vpa},
\begin{equation}
  \bm{a}^i_\mathrm{ULDM}  = -\ddot{\psi}\,\bm{r}^i.
  \label{eq:ULDMa}
\end{equation}
ULDM may also \textit{directly} couple to the constituents of the binary via non-gravitational interactions, further enhancing this acceleration~\cite{Blas:2016ddr}. In this \textit{Letter}, we extend previous studies of binaries perturbed by ULDM by applying our framework, showing that binary resonances can access new regions of ULDM parameter space.

{\bf Perturbed Orbital Dynamics.}--- %
\begin{figure}[!htb]
    \centering
    \includegraphics[width=1\linewidth]{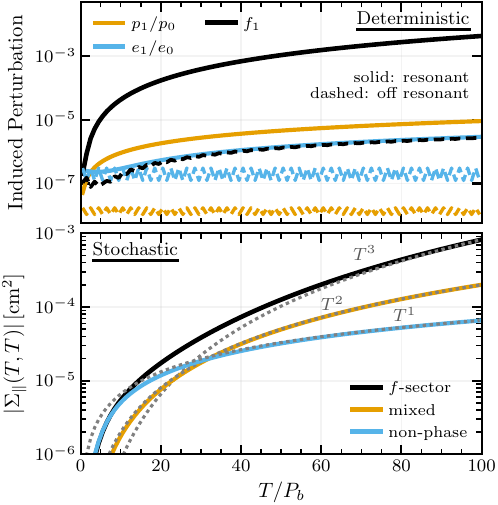}
    \caption{Illustration of time-resolved responses for the Earth--Moon system. The top panel shows representative deterministic perturbations to \(p\), \(e\), and \(f\) from monochromatic GW forcing, RMS-averaged over one orbital period. On resonance, accumulated orbital dephasing makes the true anomaly grow faster than the directly driven geometric elements; off resonance, the geometric response is oscillatory while the true anomaly can still accumulate a slower dephasing. The bottom panel shows the RMS-averaged equal-time Earth--Moon ranging covariance for an illustrative broadband stochastic GW spectrum with Gaussian \(d\rho_{\rm GW}/d\nu\), centered at \(\nu=1/P_b\), width \(0.05/P_b\), and normalization \(\rho_c^{-1}\int d\nu\, d\rho_{\rm GW}/d\nu=10^{-5}\). The covariance is decomposed as in Eq.~\eqref{eq:observable_covariance_decomposition}: the black, yellow, and blue curves show the \(f\)-sector, mixed, and geometric contributions, respectively, as defined in Eq.~\eqref{eq:broadband_scaling} and the text below.  Early transients can dominate, but the late-time envelopes recover the expected broadband hierarchy, \(T^3\), \(T^2\), and \(T\); gray dotted curves show these analytic scalings.}
    \label{fig:scaling_schematic}
\end{figure}
Working with the method of osculating elements~\cite{Poisson_Will_2014,Foster:2025csl}, the six degrees of freedom of our binary system are: the semilatus rectum $p$, the eccentricity $e$, the inclination $\iota$, the longitude of the ascending node $\Omega$, the argument of periapsis $\omega$, and the true anomaly $f$. We compactly denote these six degrees of freedom, in the order listed, as the state vector $X^\alpha\equiv\{p,e,\iota,\Omega,\omega,f\}$. These degrees of freedom evolve following a first-order system which, from Eq.~\eqref{eq:GWPerturbation}, can be expressed as~\cite{Foster:2025csl}
\begin{equation}
    \dot{X}^{\alpha} = {F}_0^{\alpha}(X, t) +  F^{\alpha}_{1}(X,t)\ddot{h}(t),
    \label{eq:EoM_Abstract}
\end{equation}
where ${F}_0^{\alpha}(X, t)$ is the total contribution of all other Newtonian and potentially non-Newtonian accelerations, and $ F^{\alpha}_{1}(X,t)$ is a transfer function capturing the effect of the GW perturbation $\ddot{h}$. We neglect the tensor indices associated with the GW for simplicity here, and encourage the interested reader to consult Ref.~\cite{Foster:2025csl} for complete expressions and further details regarding ${F}^\alpha_0$ and $F^\alpha_{1}$.
In the ULDM case, $\ddot{h}$ is simply replaced with $\ddot{\psi}$, with the corresponding replacement in \(F^{\alpha}_{1}(X,t)\).

Treating the system perturbatively in ${F}^{\alpha}_1$, we first solve for the zeroth-order evolution,
\begin{equation}
    \dot{X}_0^{\alpha} = {F}_0^{\alpha}(X_0, t),
    \label{eq:ZerothOrderEoM}
\end{equation}
which takes into account all nonlinearities in the unperturbed dynamics. 
The linearized equations of motion for the first-order term then read
\begin{equation}
\begin{gathered}
\dot{X}_1^{\alpha}  = {F}_0^{\alpha \beta}(t) X_{1}^{\beta} + {F}^{\alpha }_{1}(X_0(t), t)\ddot{h}(t),
\end{gathered}
\label{eq:FirstOrderEom_Abstract}
\end{equation}
with ${F}_0^{\alpha \beta}(t) \equiv \frac{\partial {F}_0^{\alpha }(X, t)}{\partial X^{\beta}} \big|_{X = X_0(t)}$. We solve Eq.~\eqref{eq:FirstOrderEom_Abstract} by variation of parameters using the fundamental matrix, following Ref.~\cite{Foster:2025csl}; the explicit construction is summarized in App.~\ref{app:solving}. A stationary Gaussian background yields a vanishing mean perturbation and a time-dependent covariance $\Sigma_{\alpha\beta}(t,t')$ for the osculating elements.

The expansion assumes the induced orbital perturbations remain small, with the initial orbit fixed to the unperturbed solution. Since, as shown below, the true anomaly is the fastest-growing response near resonance, we require \(\Delta f<1\) for the perturbative approach to be valid. For a monochromatic GW, \(h_{ij}(t)=h_0 e_{ij}\cos(\omega t)\), with \(e_{ij}\) a fixed polarization tensor, Eq.~\eqref{eq:GWPerturbation} gives a characteristic acceleration \(A_{\rm GW}\sim h_0\omega^2 r\), up to order-unity geometric and polarization factors. Evaluated at apocenter, \(A_{\rm GW}\sim h_0\omega^2 a_0(1+e_0)\), where \(a_0\) and \(e_0\) are the semimajor axis and eccentricity of the unperturbed orbit. The osculating evolution of the orbital period is
\begin{equation}
    \dot P
    =
    \frac{6\pi a^2}{GM}
    \frac{
        e\sin f\,\mathcal R
        +
        (1+e\cos f)\mathcal S
    }{\sqrt{1-e^2}},
    \label{eq:Pdot_osculating}
\end{equation}
where \(\mathcal R\) and \(\mathcal S\) are the radial and tangential components of the perturbing acceleration \cite{Poisson_Will_2014}. For a perturbing acceleration of magnitude \(A\), this implies the maximal dephasing estimate
\begin{equation}
    \Delta f
    \sim
    \frac{A T_\mathrm{obs}^2}{a_0}
    \frac{1+e_0}{(1-e_0)^2}.
    \label{eq:dephasing_acceleration_estimate}
\end{equation}
Inserting \(A_\mathrm{GW}\) and requiring \(\Delta f<1\) gives \(h_0\lesssim(\omega T_\mathrm{obs})^{-2}[(1-e_0)/(1+e_0)]^2\). Near the \(k\)-th orbital harmonic, \(\omega\simeq 2\pi k/P\), this becomes
\begin{equation}
    h_0
    \lesssim
    10^{-6}\,
    \left(\frac{k N_{\rm periods}}{200}\right)^{-2}
    \left(\frac{1-e_0}{1+e_0}\right)^2 ,
    \label{eq:h0_perturbativity_condition}
\end{equation}
where \(N_{\rm periods}\equiv T_\mathrm{obs}/P\). 
Using \(\Omega_{\rm GW}\sim(\omega/H_0)^2h_0^2\), the same criterion can be written as
\begin{equation}
    \Omega_{\rm GW}(\omega)
    \lesssim
     10^{11}
    \left(\frac{T_\mathrm{obs}}{10\,\mathrm{yr}}\right)^{-4}
    \left(\frac{\omega}{2\pi\,\mu\mathrm{Hz}}\right)^{-2}
    \left(\frac{1-e_0}{1+e_0}\right)^4 .
    \label{eq:omega_gw_perturbativity_condition}
\end{equation}
For the observations considered here, violating this conservative bound would require \(\Omega_{\rm GW}\gg1\), so the linearized treatment is safely within its regime of validity throughout the physically relevant parameter space.

{\bf Time-Resolved Orbital Response.}--- %
Our perturbative treatment of all six osculating elements gives a time-resolved description of the orbital response to an external acceleration. On resonance, the geometric elements \(\{p,e,\iota,\Omega,\omega\}\), which determine the shape and orientation of the orbit, can grow linearly in time. The true anomaly \(f\), which specifies the orbital phase, then plays a privileged role: the growth of \(p\) and \(e\) feeds into the Keplerian equation for \(\dot f\), sourcing a linearly growing contribution to \(\dot f_1\). Integrating once more gives \(f_1\propto T^2\). (Off resonance, the geometric elements do not grow secularly, and \(f_1\propto T\).) This quadratic scaling was studied parametrically in Ref.~\cite{Hui:2012yp} through the resonant shift in the time of periastron, but not as part of a complete time-resolved treatment of the osculating-element evolution. In contrast, secular-period treatments~\cite{Blas:2016ddr,Blas:2021mpc,Blas:2021mqw}  track the derived orbital period and do not retain this orbital-dephasing channel. These scalings, derived in detail in Ref.~\cite{Foster:2025csl}, are confirmed in the top panel of Fig.~\ref{fig:scaling_schematic} by numerically integrating the orbital perturbations in the case of the Earth-Moon system.

For stochastic signals, the same hierarchy appears at the level of the orbital-element covariance. Since a finite-time perturbation can be decomposed into Fourier modes, consider first a monochromatic SGWB component. If its frequency lies on an orbital harmonic, then the covariances of induced orbital perturbations scale as
\begin{equation}
\begin{aligned}
    \Sigma_{AB}(t,t') &\sim tt',
    &
    \Sigma_{ff}(t,t') &\sim (tt')^2,
    \\
    \Sigma_{Af}(t,t') &\sim t(t')^2,
    &
    \Sigma_{fA}(t,t') &\sim t^2t',
\end{aligned}
    \label{eq:monochromatic_cov_scaling}
\end{equation}
where \(A,B\in\{p,e,\iota,\Omega,\omega\}\). Hence, at equal times $t = t' = T$, the geometric, mixed, and true-anomaly blocks scale as \(T^2\), \(T^3\), and \(T^4\), respectively. Off resonance, the loss of secular growth in the geometric elements reduces these equal-time scalings by \(T^2\): the geometric block remains bounded, the mixed blocks scale as \(T\), and the true-anomaly block scales as \(T^2\).

The broadband SGWB response is obtained by integrating the monochromatic covariance over frequency, as in Eq.~\eqref{eq:covariancefield} in App.~\ref{app:solving}. If the relevant harmonic \(\nu_k=k/P_b\) is effectively constant over the observation time \(T\), only modes within the finite-time resonant bandwidth \(\Delta\nu_{\rm res}\sim1/T\) add coherently. For a smooth spectrum \(P_h(\nu)\), the coherent monochromatic scaling is therefore softened by one power of \(T\):
\begin{equation}
    \Sigma_{ff}^{\rm SGWB}(T,T)
    \sim
    P_h(\nu_k)\,T^4\,\Delta\nu_{\rm res}
    \sim
    P_h(\nu_k)\,T^3 ,
    \label{eq:broadband_scaling}
\end{equation}
up to factors of \(\nu_k\) and orbit-dependent transfer functions. If the harmonic drifts appreciably, the same argument applies on coherent intervals \(\tau_j\) satisfying \( |\dot\nu_k|\tau_j^2\lesssim1 \), with local bandwidth \(\Delta\nu_j\sim1/\tau_j\):
\begin{equation}
    \Sigma_{ff}^{\rm SGWB}
    \sim
    \sum_j
    P_h(\nu_k(t_j))\,\tau_j^4\,\Delta\nu_j
    \sim
    P_h(\nu_k)\,T\,\tau^2 ,
    \label{eq:broadband_drift_scaling}
\end{equation}
where the second scaling assumes a slowly varying spectrum and characteristic coherent time \(\tau\) over observing time \(T\). In the constant-harmonic case the geometric, mixed, and true-anomaly blocks scale as \(P_hT\), \(P_hT^2\), and \(P_hT^3\); with drift they scale as \(P_hT\), \(P_hT\tau\), and \(P_hT\tau^2\). Thus the true-anomaly block is the most rapidly growing contribution to the orbital-element covariance.

\begin{figure*}[!htb]
\includegraphics[width=.95\textwidth]{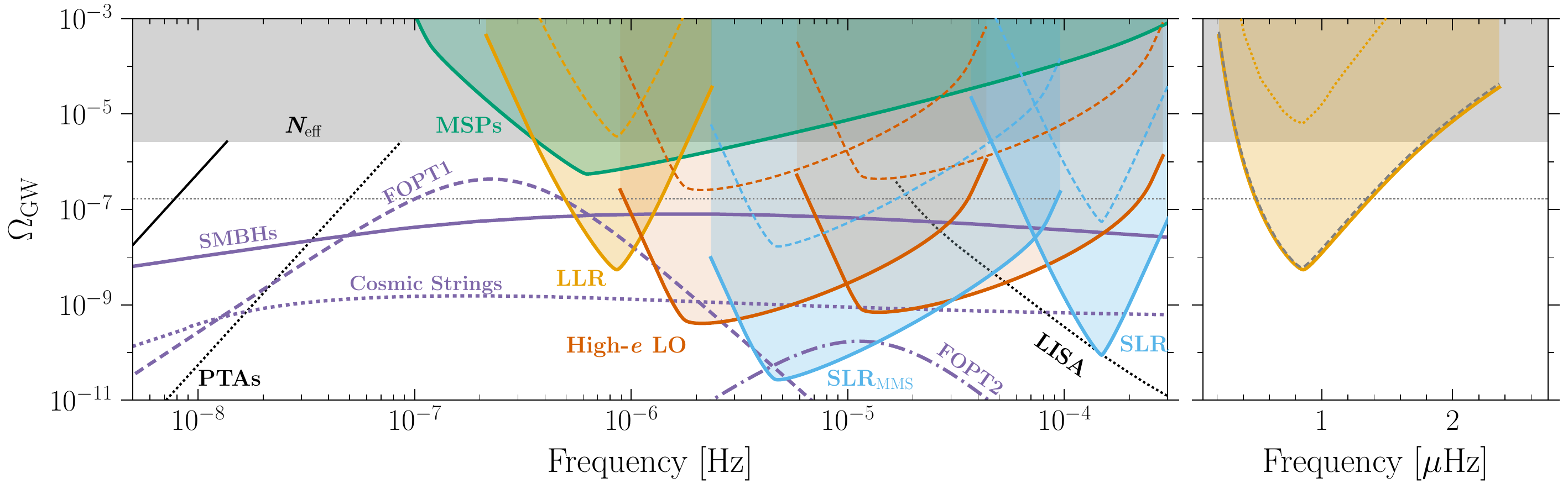}
\caption{(\textit{Left}) Projected power-law-integrated sensitivities of observations of binary systems  to power-law SGWBs for each of the measurement scenarios discussed in this work, at frequencies between $10^{-8}\,\mathrm{Hz}$ and $10^{-4}\,\mathrm{Hz}$. We show Lunar Laser Ranging (LLR), Satellite Laser Ranging (SLR), Laser Ranging of a Magnetospheric Multiscale Mission-like Satellite ($\mathrm{SLR}_\mathrm{MMS}$), Milli-Second Pulsars (MSP), and two possibilities for highly eccentric lunar orbits (high-$e$ LO) and consider two scenarios: \textit{optimistic} (solid) and \textit{pessimistic} (dashed), see text for more details. We also depict: current and future $N_\mathrm{eff}$ sensitivities in solid grey and dotted grey, respectively~\cite{Pagano:2015hma}; current and future pulsar timing array sensitivities (PTAs)~\cite{Janssen:2014dka,Lasky:2015lej, NANOGrav:2023gor} in solid black and dotted black, respectively; and the future reach of LISA~\cite{amaroseoane2017laserinterferometerspaceantenna} in dotted black. Finally, we show possible signals from supermassive black hole binaries (SMBH), two first-order phase transitions showing the complementarity with PTA and LISA (FOPT1 and FOPT2) and cosmic strings in solid purple, dashed purple, and dotted purple, respectively. (\textit{Right}) Comparison of the projected LLR sensitivity obtained with our time-resolved framework to the secular-period projections of Refs.~\cite{Blas:2021mpc,Blas:2021mqw}, assuming identical LLR precision. The dashed grey curve shows the projected sensitivity when the solar tidal perturbation is included, demonstrating the robustness of the resonant response.}
\label{fig:GWReach}
\vspace{-2ex}
\end{figure*}

Cold ULDM qualitatively differs from a broadband SGWB because its fractional linewidth is set by the Galactic velocity dispersion, $\Delta\nu_{\rm ULDM}/\nu_{\rm ULDM}\sim v^2\sim10^{-6}$. Resolving this linewidth near the $k$th harmonic would require \(T_{\rm obs}\Delta\nu_{\rm ULDM}\sim k v^2T_{\rm obs}/P_b\gtrsim1\), which is not satisfied for the systems and harmonics relevant to our forecasts. The ULDM perturbation is therefore effectively monochromatic over the measurement campaign: if it lies within the finite-time resonant band, \(\Sigma_{ff}^{\rm ULDM}\sim(tt')^2\), with associated slower growth in the mixed and geometric covariance blocks; otherwise it behaves as an off-resonance component, with bounded geometric-element covariance and only slower linear dephasing in \(f\).

While these asymptotic scalings guide the interpretation, our forecasts use the full frequency-dependent response. For the systems considered here, coherence times computed at the first orbital harmonic are long compared with the projected observing times: using the observed lunar recession rate and LAGEOS semimajor-axis decay gives coherence times of approximately \(10^4\,{\rm yr}\) for the Earth--Moon system, $10^2 \,\mathrm{yr}$ for LAGEOS, and \(\gtrsim10^5\,{\rm yr}\) for representative pulsars. Since any non-Keplerian accelerations which act on these systems do not cause them to detune on the relevant observational timescales, we approximate them as stable Keplerian systems. 

{\bf Targets, Observables, and Sensitivity.}--- %
We consider four classes of precision-tracked binaries; see App.~\ref{app:data} for target details. LLR measures the Earth--Moon separation through the two-way light-travel time $\Delta_{||}$, SLR provides analogous Earth--satellite observables, and binary millisecond pulsars provide one-way timing observables $\Delta_{|}$. We also consider high-eccentricity artificial-satellite scenarios motivated by Moonlight and Gateway~\cite{esaESAsMoonlight,nasaGatewayNASA}. Our forecasts use selected systems whose orbital evolution is well-modeled over the observing time and for which the GW/ULDM-induced perturbations remain in the linear regime.

For small perturbations, both $\Delta_{|}$ and $\Delta_{||}$ can be expanded to first order in the osculating-element perturbations,
\begin{equation}
    \Delta_{i}\approx X^{\alpha}_1 
    \frac{\partial \Delta_i}{\partial X^{\alpha}}   
    \bigg|_{X=X_0(t)}
    \equiv {T}_{i\alpha}(t) X^{\alpha}_1,
\end{equation}
where \(i=\{|,||\}\), and \(T_{i\alpha}\) maps orbital-element perturbations into timing or ranging residuals; see App.~\ref{app:PerturbationMapping}. For a Gaussian isotropic SGWB, the mean first-order residual vanishes, while the covariance is
\begin{equation}
\begin{gathered}
\Sigma_i(t, t') 
= \langle \Delta_i(t)\Delta_i(t') \rangle 
= {T}_{i\alpha}(t) {T}_{i\beta}(t') {\Sigma}_{\alpha\beta}(t, t').
\end{gathered}
\label{eq:mu_sigma}
\end{equation}
Separating the true anomaly from the geometric elements \(A,B\in\{p,e,\iota,\Omega,\omega\}\), we can write
\begin{equation}
\begin{aligned}
\Sigma_i(t,t') ={}&
T_{if}(t)T_{if}(t')\Sigma_{ff}
+T_{if}(t)T_{iA}(t')\Sigma_{fA} \\
+&T_{iA}(t)T_{if}(t')\Sigma_{Af}
+T_{iA}(t)T_{iB}(t')\Sigma_{AB} ,
\end{aligned}
\label{eq:observable_covariance_decomposition}
\end{equation}
where all covariances are evaluated at \((t,t')\). 

Because the transfer functions $T_{i\alpha}$ are orbital-periodic, the late-time hierarchy is inherited from the orbital covariances: the \(f\)-sector grows fastest, the mixed blocks are intermediate, and the geometric block grows slowest. Retaining the \(f\)-sector is therefore the parametric origin of the improvement over secular-period projections~\cite{Blas:2021mpc,Blas:2021mqw}, which track derived quantities such as the orbital period and discard the body's accumulated orbital phase. This effect is illustrated in the bottom panel of Fig.~\ref{fig:scaling_schematic}. Relative to Ref.~\cite{Hui:2012yp}, the quadratic dephasing is incorporated here as a modeled contribution to the full timing or ranging covariance, rather than only through a derived shift in the time of periastron, allowing the likelihood to preserve the information carried by the full residual time series.

Our framework therefore allows us to write the likelihood at the level of the data, rather than in terms of derived orbital quantities. Assuming Gaussian white measurement noise, the likelihood of a data series $\bm d$ is
\begin{equation}
    \mathcal{L}(\bm{d} \mid \bm{\theta})
    = \frac{\exp\Bigl[
        -\bigl(\bm{d} - \bm{\mu}(\bm{\theta})\bigr)^{\mathsf{T}}
         \bigl(\sigma^{2} \bm{I} + \bm{\Sigma}_{i}\bigr)^{-1}
         \bigl(\bm{d} - \bm{\mu}(\bm{\theta})\bigr) / 2
      \Bigr]}
     {\Bigl\lvert 2 \pi \bigl(\sigma^{2} \bm{I} + \bm{\Sigma}_{i}\bigr) \Bigr\rvert^{1/2}},
\end{equation}
where $\bm{\mu}$ is the time-resolved orbital model prediction in the absence of a SGWB or ULDM. 
Writing the signal covariance as $\bm\Sigma_i=\mathcal{A} \bm S_i$, where $\mathcal{A}$ is the background amplitude and $\bm S_i$ is the canonically normalized covariance, the weak-signal sensitivity to $\mathcal{A}$ is controlled by the one-parameter Fisher information,
$F_{\mathcal{A}\mathcal{A}}\equiv-\langle \partial_{\mathcal{A}}^2\ln\mathcal{L}\rangle$~\cite{Vallisneri:2007ev}. Evaluating this under the null hypothesis gives
\begin{equation}
    F_{\mathcal{A}\mathcal{A}}
    =
    \frac{1}{2\sigma^4}
    \mathrm{Tr}\!\left[\bm S_i^2\right].
    \label{eq:weak_signal_fisher}
\end{equation}
Thus the detectable amplitude scales inversely with the covariance norm. Retaining the true-anomaly block therefore enhances sensitivity over secular-period treatments by a factor scaling as $(T_{\rm obs}/P_b)^2$, with an eccentricity-dependent prefactor determined by the orbital response and its mapping to the observable~\cite{Foster:2025csl}.

\begin{figure}[!tb]
\includegraphics[width=.48\textwidth]{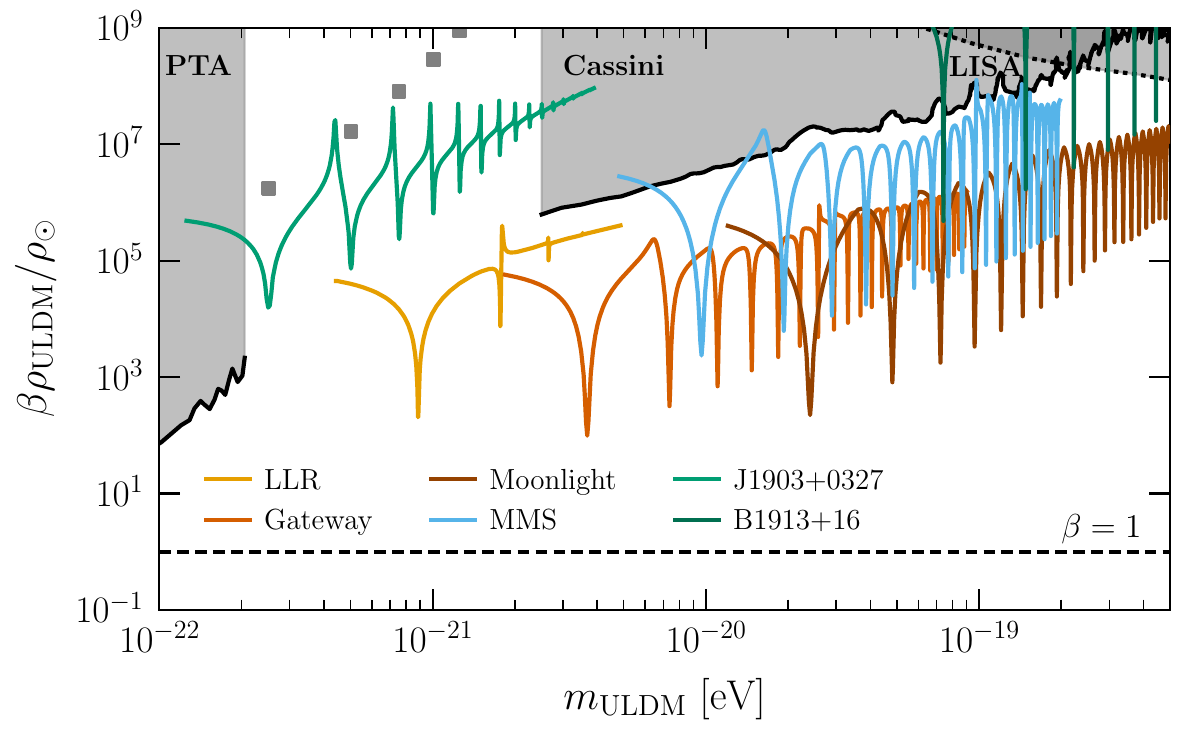}
\caption{Projections, as in Fig.~\ref{fig:GWReach}, but for the sensitivities of the binary resonance scenarios considered in this work to ultra-light dark matter. Grey squares indicate the projected on-resonance sensitivity for J1903+0327 from Refs.~\cite{Blas:2016ddr,Blas:2019hxz}, upon which our projections improve considerably.}
\label{fig:ULDMReach}
\vspace{-2ex}
\end{figure}

{\bf GW and ULDM sensitivities.}--- %
For each observational target, we project sensitivities to SGWBs in Fig.~\ref{fig:GWReach} and to ULDM in Fig.~\ref{fig:ULDMReach}; deterministic GW signals are treated in Ref.~\cite{Foster:2025csl}. Our optimistic scenario adopts the ``2021 Sensitivity'' measurement precisions of Refs.~\cite{Blas:2021mpc,Blas:2021mqw}, consistent with present timing and laser-ranging capabilities, using the statistical noise described in App.~\ref{app:data}. To bracket systematics and modeling uncertainties, the pessimistic scenario inflates the effective error budget by a factor of 100. This is motivated by the gap between raw LLR ranging precision and current residual modeling accuracy~\cite{muller:2019tv,*park:2021wd,bourgoin:2017aa,*bourgoin:2021wu}, with analogous considerations for SLR and pulsar timing~\cite{lucchesi:2014ph,Taylor:2021yjx}.

We project sensitivity to the SGWB energy density spectrum, $\Omega_{\rm GW}(\nu)$, using the power-law-integrated method~\cite{Thrane:2013oya}; see App.~\ref{app:PLIntegrated}. Figure~\ref{fig:GWReach} also shows representative $\mu$Hz spectra from SMBH binaries consistent with the PTA signal~\cite{Ellis:2023owy}, first-order phase transitions~\cite{Caprini:2018mtu}, and cosmic strings~\cite{Renzini:2022alw}. Including the solar gravitational potential, the dominant perturbation of the Earth--Moon system, leaves its projected sensitivity essentially unchanged~\cite{Foster:2025csl}. For LLR, $T_{\rm obs}/P_b\simeq200$, so the quadratic scaling is parametrically consistent with an enhancement of several orders of magnitude; the full calculation yields the $\sim1.2\times10^3$ improvement over secular-period projections shown in the right panel of Fig.~\ref{fig:GWReach}.

We also project sensitivity to ULDM-induced oscillations of the binary potential. We consider gravitational interactions and universal quadratic direct couplings, though the method generalizes to linear and nonuniversal couplings~\cite{Blas:2016ddr,Blas:2019hxz}. The gravitational effect is described by Eq.~\eqref{eq:ULDMa}; a universal quadratic coupling gives $\ddot\psi\rightarrow\beta\ddot\psi$, with $\beta\equiv\pi G_N\Lambda^2$. We use the near-monochromatic component at $2\pi\nu=2m_{\rm ULDM}$ and express the reach in terms of $\beta\rho_{\rm ULDM}$, with $\beta=1$ corresponding to gravity alone. Fig.~\ref{fig:ULDMReach} shows that selected targets can improve over Cassini bounds~\cite{Bertotti:2003rm} and anticipated LISA sensitivity~\cite{Kim:2023pkx} while also opening access to ULDM masses that would otherwise remain unprobed.

{\bf Discussion.}--- %
In this \textit{Letter}, we have demonstrated that GW searches via the binary resonance mechanism \emph{could achieve leading sensitivity across as many as four decades in frequency}. The same probes can also access the gravitational effects of ULDM in dark-matter overdensities, as well as non-gravitational interactions between ULDM and matter. This sensitivity would impact searches for GWs from SMBH binaries, shed light on the formation and evolution of SMBHs and their host galaxies, and probe new physics generating GWs in the early Universe~\cite{Blas:2021mqw,Sesana:2019vho}. Our results also motivate incorporating GW/ULDM sensitivity into the design and commissioning of upcoming satellites and satellite networks. Most notably, current LLR data may already be sensitive to several well-motivated GW sources; we plan to explore this possibility in future work.

{\bf Acknowledgements.}---%
{\it We thank D.~Hackett, Y.~Kahn, N.~Rodd, B.~Safdi, T.~Trickle, and J.~Urban for helpful discussions. This research used resources from the Lawrencium computational cluster provided by the IT Division at the Lawrence Berkeley National Laboratory, supported by the Director, Office of Science, and Office of Basic Energy Sciences, of the U.S. Department of Energy under Contract No. DE-AC02-05CH11231.
Part of this work was carried out at the Munich Institute for Astro-, Particle and BioPhysics (MIAPbP), which is funded by the Deutsche Forschungsgemeinschaft (DFG, German Research Foundation) under Germany's Excellence Strategy – EXC-2094 – 390783311. This manuscript has been authored in part by Fermi Forward Discovery Group, LLC under Contract No. 89243024CSC000002 with the U.S. Department of Energy, Office of Science, Office of High Energy Physics.

D.B. acknowledges the support from the Departament de Recerca i Universitats from Generalitat de Catalunya to the Grup de Recerca 00649 (Codi: 2021 SGR 00649).
The research leading to these results has received funding from the Spanish Ministry of Science and Innovation (PID2020-115845GB-I00/AEI/10.13039/501100011033). This publication is part of the grant PID2023-146686NB-C31 funded by MICIU/AEI/10.13039/501100011033/ and by FEDER, UE.
IFAE is partially funded by the CERCA program of the Generalitat de Catalunya. Project supported by a 2024 Leonardo Grant for Scientific Research and Cultural Creation from the BBVA Foundation. The BBVA Foundation accepts no responsibility for the opinions, statements and contents included in the project and/or the results thereof, which are entirely the responsibility of the authors. This work is supported by ERC grant ERC-2024-SYG 101167211. Funded by the European Union.  Views and opinions expressed are however those of the author(s) only and do not necessarily reflect those of the European Union or the European Research Council Executive Agency. Neither the European Union nor the granting authority can be held responsible for them. D.B. acknowledges the support from the European Research Area (ERA)
via the UNDARK project of the Widening participation
and spreading excellence programme (project number 101159929). 

The work of M. H-V. has been partially supported by the Spanish State Research Agency MCIN/AEI/10.13039/501100011033 and the EU NextGenerationEU/PRTR funds, under grant IJC2020-045126-I.

A.C.J. was supported by a Gavin Boyle Fellowship at the Kavli Institute for Cosmology, Cambridge, and by the Science and Technology Facilities Council through the UKRI Quantum Technologies for Fundamental Physics Programme (Grant No. ST/T005904/1).

X.X. is funded by the grant CNS2023-143767. 
Grant CNS2023-143767 funded by MICIU/AEI/10.13039/501100011033 and by European Union NextGenerationEU/PRTR.}

\bibliography{refs}

\clearpage

\appendix

\section{Time-resolved solutions}
\label{app:solving}

Here we summarize the time-resolved solution of the linearized orbital perturbations used in the main text. The goal is to obtain the mean and covariance of the first-order perturbations $X_1^\alpha$ induced by a stationary stochastic background.

We begin from the homogeneous system
\begin{equation}
\dot{X}_1^{\alpha}  = {F}_0^{\alpha\beta}(t) X_{1}^{\beta},
\end{equation}
and define the fundamental matrix $\Phi_{\alpha\beta}(t)$ by $X_1^\alpha(t)=\Phi_{\alpha\beta}(t)X_1^\beta(0)$. It satisfies
\begin{equation}
    \dot{\Phi}_{\alpha\beta} = {F}_{0}^{\alpha\kappa}\Phi_{\kappa \beta},
    \qquad
    \Phi_{\alpha\beta}(0) = \delta_{\alpha\beta}.
\label{eq:StateTransitionEOM}
\end{equation}
The solution to Eq.~\eqref{eq:FirstOrderEom_Abstract} with $X^\alpha_1(0)=0$ is then
\begin{equation}
    X^{\alpha}_1(t) =
    \Phi_{\alpha\beta}(t)
    \int_0^t \mathrm{d}\tau \,
    \Phi^{-1}_{\beta\gamma}(\tau)\,
    {F}^{\gamma}_{1}(\tau)\ddot{h}(\tau),
    \label{eq:ParticularSolution}
\end{equation}
where $\Phi^{-1}$ is the inverse of the fundamental matrix.

For a stationary, Gaussian, isotropic stochastic background in the center-of-mass frame, the orbital perturbations have vanishing mean $\mu_\alpha(t) = 0$ and covariance
\begin{equation}
\begin{aligned}
&\Sigma_{\alpha\beta}(t, t')\equiv \langle X_1^\alpha(t) X_1^\beta(t')\rangle 
= \Phi_{\alpha \gamma}(t) \Phi_{\beta \lambda}\left(t^{\prime}\right)\\
&\quad\times\int_0^\infty \mathrm{d}\nu \,\nu P_h(\nu) [C_\gamma(t,\nu)C_\lambda(t',\nu)+S_\gamma(t,\nu)S_\lambda(t',\nu)].
\end{aligned}
\label{eq:element_covariance}
\end{equation} 
where
\begin{equation}
\langle \ddot{h}(t)\ddot{h}(t') \rangle
=
\int_0^\infty \mathrm{d}\nu\, \nu P_h(\nu)\cos[2\pi\nu(t-t')].
\label{eq:covariancefield}
\end{equation}
Here $P_h(\nu)$ denotes the appropriate perturbation power spectrum; see Ref.~\cite{Foster:2025csl} for the corresponding expressions for GWs and ULDM, including the tensor structure in the GW case. The auxiliary functions $C_\alpha$ and $S_\alpha$ satisfy
\begin{equation}
\begin{aligned}
\dot{C}_{\alpha}(t,\nu) &= \Phi^{-1}_{\alpha\beta}(t){F}^\beta_{1}(t)\cos(2\pi \nu t), \\
\dot{S}_{\alpha}(t,\nu) &= \Phi^{-1}_{\alpha\beta}(t){F}^\beta_{1}(t)\sin(2\pi \nu t).
\end{aligned}
\end{equation}
Together, the vanishing mean and Eq.~\eqref{eq:element_covariance} define the Gaussian statistics of the time-resolved perturbations to the osculating elements and provide the covariance input used in the main text.
\section{Data sets}
\label{app:data}

Here we further specify potential observational targets and associated assumed sensitivities used to develop the forecasts in Fig.~\ref{fig:GWReach} and Fig.~\ref{fig:ULDMReach}. In each case, we assume that the measurement points are uniformly distributed in time over the orbit.

{\it Lunar Laser Ranging.}-- %
The APOLLO collaboration has performed LLR measurements since 2006, reporting roughly 260 normal points per year with $3\,\mathrm{mm}$ uncertainty per measurement~\cite{Murphy:2013qya}. Following~\cite{Blas:2021mpc,Blas:2021mqw}, we adopt 15 years of observations at this precision as our fiducial ``2021 sensitivity'' scenario, allowing direct comparison with earlier forecasts~\cite{Blas:2021mpc,Blas:2021mqw,Blas:2016ddr,Blas:2019hxz,Kus:2024vpa}. An additional $\sim 40$ years of LLR data exist at lower precision. Current orbital models reach residuals of a few cm rather than the raw mm ranging precision~\cite{muller:2019tv,*park:2021wd}, motivating the pessimistic scenario discussed in the main text. However, future NGLR deployments through CLPS, Artemis, the Lunar Geophysical Network, and Moonlight are expected to improve both ranging precision and lunar geophysical modeling, including tidal dissipation and moment-of-inertia constraints~\cite{2022PSJ.....3..136W}.

Additionally, as we demonstrate in~\cite{Foster:2025csl}, resonant accelerations result in light-travel-time perturbations which grow quadratically in time, while non-resonant accelerations grow only linearly. This allows for disentangling a resonant signal from non-resonant mismodeling. In a preliminary Fisher analysis we have found that the GW signal decorrelates well from most fitted parameters in LLR data analysis. An exception is the tidal interaction, which also generates a resonant acceleration, but can be described by a model specified by independently inferred parameters \cite{esaESAsMoonlight}.

{\it Satellite Laser Ranging: existing missions.}-- %
Following~\cite{Blas:2021mpc,Blas:2021mqw}, we use the LAGEOS1 satellite as our representative present-day SLR target. LAGEOS1 has orbited Earth since 1976 on an orbit with $e=0.0045$ and period $\sim4\,\mathrm{hours}$~\cite{Ciufolini:2016ntr}. We assume 45 years of observations, with 50,000 measurements per year at $3\,\mathrm{mm}$ precision~\cite{Ciufolini:2016ntr}.

{\it Satellite Laser Ranging: future missions.}-- %
As compared to LLR, the sensitivity from SLR can be dramatically increased with respect to the LAGEOS mission in two ways: by considering SLR of highly eccentric orbits and by tracking a constellation of optimized orbits, cf.~\cite{Foster:2025csl}. To investigate these possibilities, we consider the laser ranging of satellites of both Earth-based and lunar-based selected orbits. For the Earth-based case, we consider the orbit of the MMS mission at Phase 2b \cite{MMSNASA}, with $e=0.9084 $ and a perigee altitude of 2,550 km. This orbit will not be affected by the Earth's atmosphere, but laser tracking at these large distances and the existence of other noise sources, see, \textit{e.g.}, \cite{app122110772}, requires further study. We also consider two proposed high-eccentricity orbits around the Moon. The first is the Gateway orbit at a period of approximately 6.5 days and an eccentricity of 0.7~\cite{nasaGatewayNASA}; the second is the Moonlight constellation, consisting of four or more satellites on unaligned orbits with a period of one day and an eccentricity of $0.6$~\cite{esaESAsMoonlight}. For each scenario, we assume a 10-year data-taking campaign with a laser-ranging uncertainty 10 times greater than the ``present'' uncertainty assumed for the LAGEOS1 projection.

These examples illustrate the reach enabled by high eccentricity and optimized satellite constellations. Although the specific missions considered here may not be ideal GW/ULDM targets, the required orbit placement and high-precision ranging technology are within reach.

\begin{table}[!t]
\centering
\begin{tabular}{|c|c|c|c|c|c|c|}
\hline
Binary Pulsars       & $P$ & $e$ & $\iota$  & $\omega$ & $M_1$  & $M_2$  \\ \hline \hline
J0737-3039A~\cite{Burgay:2004vz}   & 0.102   & 0.0878   & 1.55   & 3.57   & 1.34   & 1.25   \\ \hline 
B1913+16~\cite{Weisberg:2004hi}    & 0.323   & 0.617    & 0.822  & 5.11   & 1.44   & 1.39   \\ \hline
B2127+11C~\cite{Jacoby:2006dy}     & 0.335   & 0.681    & 1.05*  & 6.03   & 1.36   & 1.35   \\ \hline
B1534+12~\cite{Fonseca:2014qla}    & 0.420   & 0.274    & 1.35   & 4.95   & 1.33   & 1.35   \\ \hline
J1829+2456~\cite{Haniewicz:2020jro} & 1.18    & 0.139    & 1.32   & 4.01   & 1.31   & 1.57   \\ \hline
B2303+46~\cite{Kerkwijk:1999xj}    & 12.3    & 0.658    & 1.05*  & 0.612  & 1.16   & 1.37   \\ \hline
J0514-4002A~\cite{Ridolfi:2019wgs} & 18.8    & 0.888    & 0.908  & 1.44   & 1.25   & 1.22   \\ \hline
J1748-2021B~\cite{Freire:2007jd, Clifford2019} & 20.6 & 0.570    & 1.05*  & 5.49   & 2.55   & 0.0790 \\ \hline
J1946+3417~\cite{Barr:2016vxv}     & 27.0    & 0.135    & 1.33   & 3.90   & 1.83   & 0.266  \\ \hline
J1903+0327~\cite{Champion:2008ge, 2011MNRAS.412.2763F} & 95.2 & 0.437 & 1.35   & 2.47   & 1.67   & 1.03   \\ \hline
\end{tabular}
\caption{Table of masses and orbital elements of the binary MSP systems considered in this work. Masses are provided in units of $M_\odot$ and the period $P$ is provided in units of days.}
\label{tab:binary_systems}
\vspace{-.4cm}
\end{table}

{\it Millisecond Pulsar Binaries.}-- %
We consider 10 MSPs in binary systems as observational targets, summarized in Tab.~\ref{tab:binary_systems}. We mark systems for which the inclination is presently unknown with an asterisk (*), and assume $\iota = \pi/3$ as in~\cite{Blas:2021mpc, Blas:2021mqw}.  The sensitivity scenario we adopt from~\cite{Blas:2021mpc, Blas:2021mqw} is that of a biweekly measurement of pulse times of arrival, with a measurement precision of $1\,\mu\mathrm{s}$ in each session, both within reach of current experimental capacity~\cite{EPTA:2023sfo,NANOGrav:2023hde}. For uniformity, we assume each pulsar is timed at this sensitivity for 15 years. Next-generation radio telescopes such as SKA and DSA-2000 should also improve on this timing scenario.

\section{Perturbation Mappings}
\label{app:PerturbationMapping}
Here we summarize the mappings from the perturbed orbital elements to the observable timing delays used in the main text.

For laser ranging, assuming emission and reception at the center of mass, the light-travel time is given by 
\begin{equation}
    \Delta_{||}(\tuple) =  \frac{2 p}{c(1+e\cos f)} +\Delta^0_{||},
    \label{eq:twoways}
\end{equation}
where $\Delta^0_{i}$ represents the noise of the measurement and also other parameters beyond the leading treatment in the $i^{\rm th}$ observable. For pulsar timing, we adopt a simplified timing model 
\begin{equation}
    \Delta_{|}=\Delta_{R}+\Delta_{S}+\Delta_{E} +\Delta^0_{|},
    \label{eq:oneway}
\end{equation}
with
\begin{equation}
\begin{aligned}
\Delta_{R} &= \frac{M_2 p \sin \iota}{c (1 - e^2) (M_1 + M_2)} 
\big[ (\cos E - e) \sin \omega  \\
&\quad  + \sqrt{1 - e^2} \sin E \cos \omega \big], \\
\Delta_{S} &= - \frac{2 G M_2}{c^3} 
\log \big[ 1 - e \cos E - \sin \iota \big( \sin \omega (\cos E - e)  \\
&\quad   + \sqrt{1 - e^2} \cos \omega \sin E \big) \big], \\
\Delta_{E} &= \left( \frac{G e^2 p}{1 - e^2} \right)^{1/2} 
\frac{M_2 (M_1 + 2 M_2)}{(M_1 + M_2)^{3/2} c^2} \sin E,
\end{aligned}
\end{equation}
which are the R\"omer, Shapiro, and Einstein time delays, respectively. Here $M_1$ is the mass of the pulsar, $M_2$ is the mass of the companion, and $E$ is the eccentric anomaly, calculable from the true anomaly $f$~\cite{Maggiore:1900zz}.

By contrast, the period-based sensitivities of Refs.~\cite{Blas:2021mpc,Blas:2021mqw} are based on
\begin{equation}
    P = \frac{2\pi}{\sqrt{GM}}\left(\frac{p}{1-e^2}\right)^{3/2},
\end{equation}
where $M$ is the total mass of the binary. This treatment does not retain the dependence of the observable time delay on the true anomaly and therefore misses the dominant contribution near resonance.

\section{Power-Law Integrated Sensitivities}
\label{app:PLIntegrated}
Here we summarize the construction of the projected sensitivities shown in Fig.~\ref{fig:GWReach}.  We consider power-law backgrounds of the form
\begin{equation}
    \Omega_{\rm GW}(\nu) = \Omega_{\rm ref}\left(\frac{\nu}{\nu_{\rm ref}}\right)^{\alpha},
    \label{eq:OmegaPowerLaw}
\end{equation}
where $\alpha$ is the spectral index and $\nu_{\rm ref}$ is a reference frequency.
For a stationary Gaussian background, the effect of the SGWB enters our time-domain likelihood only through the signal covariance $\Sigma_i$ of the observable (cf.\ Eq.~\eqref{eq:mu_sigma}), which is linear in $\Omega_{\rm GW}$ and therefore in $\Omega_{\rm ref}$ for fixed $\alpha$. We thus write
\begin{equation}
    \bm \Sigma_i(\Omega_{\rm ref},\alpha)=\Omega_{\rm ref}\, \bm S_i(\alpha),
\end{equation}
where $S_i(\alpha)$ is the covariance template obtained by inserting Eq.~\eqref{eq:OmegaPowerLaw} into Eq.~\eqref{eq:mu_sigma}.
Assuming white measurement noise with variance $\sigma^2$, the total covariance is
$C=\sigma^2 \bm I+\bm \Sigma_i$.
We quantify the sensitivity to $\Omega_{\rm ref}$ using a Fisher forecast under the null hypothesis, yielding
\begin{equation}
    \sigma_{\Omega_{\rm ref}}(\alpha)
    =\left[\frac{1}{2\sigma^4}\mathrm{Tr}\!\left(\bm S_i(\alpha)^2\right)\right]^{-1/2}.
\end{equation}
Setting the threshold signal-to-noise ratio to unity gives
\begin{equation}
    \Omega_{\rm ref}^{\rm min}(\alpha)=\sigma_{\Omega_{\rm ref}}(\alpha),
\end{equation}
which implies a frequency-dependent sensitivity for that power law,
\begin{equation}
    \Omega_{\rm GW}^{\rm min}(\nu;\alpha)
    =\Omega_{\rm ref}^{\rm min}(\alpha)\left(\frac{\nu}{\nu_{\rm ref}}\right)^{\alpha}.
\end{equation}
Finally, following the power-law integrated procedure of Ref.~\cite{Thrane:2013oya},
we take the plotted envelope to be
\begin{equation}
    \Omega_{\rm PLI}(\nu)=\max_{\alpha\in[\alpha_{\rm min},\alpha_{\rm max}]}\,
    \Omega_{\rm GW}^{\rm min}(\nu;\alpha),
\end{equation}
with $\alpha_{\rm min}=-10$ and $\alpha_{\rm max}=10$.

\end{document}